\documentclass[10pt, conference, letterpaper]{IEEEtran}
\IEEEoverridecommandlockouts
\usepackage{cite}
\usepackage{amsmath,amssymb,amsfonts}
\usepackage{algorithmic}
\usepackage{graphicx}
\usepackage{textcomp}
\usepackage[dvipsnames,svgnames]{xcolor}
\def\BibTeX{{\rm B\kern-.05em{\sc i\kern-.025em b}\kern-.08em
    T\kern-.1667em\lower.7ex\hbox{E}\kern-.125emX}}

\usepackage{balance}

\usepackage{multirow}
\usepackage{booktabs}
\usepackage{tabularx}
\usepackage{arydshln}

\usepackage{caption}
\usepackage[position=b]{subcaption}

\usepackage{dblfloatfix} 
\graphicspath{{../figs/}}

\usepackage{url}

\usepackage{tikz}
\usetikzlibrary{calc,arrows,arrows.meta,external,colorbrewer,matrix,shapes,shapes.misc,positioning,backgrounds,fit,math,colorbrewer,shapes.geometric,patterns,patterns.meta}
\usepackage{pgfplots}
\usepackage{pgfplotstable}
\usepgfplotslibrary{colorbrewer}
\pgfplotsset{compat=1.18}

\pgfplotscreateplotcyclelist{caaColorList}{
    {Set2-E}, %
    {Set2-B}, %
    {Set2-C}, %
    {Set2-D}, %
    {Set2-H}  %
}

\usepackage{hyperref}

\makeatletter
\def\sectionautorefname{\S\@gobble}
\def\subsectionautorefname{\S\@gobble}
\def\subsubsectionautorefname{\S\@gobble}
\makeatother

\newcommand{\result}[1]{}

\definecolor{myred}{cmyk}{0, 0.7808, 0.4429, 0.1412}

\newcommand{\done}[1]{}

\definecolor{positive}{HTML}{3C8031}
\definecolor{negative}{HTML}{AF3235} %
\definecolor{neutral}{HTML}{E1A205} %

\usepackage{pifont}

\usepackage{xspace}
\newcommand{\etal}{\textit{et al.}~}
\newcommand{\eg}{\textit{e.g.,}~}
\newcommand{\ie}{\textit{i.e.,}~}
\newcommand{\etc}{\textit{etc.}\xspace}
\newcommand{\one}{({\em i})\xspace}
\newcommand{\two}{({\em ii})\xspace}
\newcommand{\three}{({\em iii})\xspace}

\makeatletter
\renewcommand{\paragraph}[1]{\vspace*{0.03in}\noindent{\bf #1.}\hspace{0.25ex \@plus1ex \@minus.2ex}}
\newcommand{\paragraphNoDot}[1]{\vspace*{0.03in}\noindent{\bf #1}\hspace{0.25ex \@plus1ex \@minus.2ex}}
\makeatother

\newcommand*\dhline{\specialrule{0pt}{1pt}{0pt}\hdashline[.4pt/3pt]\specialrule{0pt}{0pt}{2pt}}
\newcommand{\dcline}[1]{\specialrule{0pt}{1pt}{0pt}\cdashline{#1}[.4pt/3pt]\specialrule{0pt}{0pt}{2pt}}

\IEEEoverridecommandlockouts

\begin{document}
\bstctlcite{IEEEexample:BSTcontrolNew}

\title{Do CAA, CT, and DANE Interlink in Certificate Deployments? A Web PKI Measurement Study}

\author{
  \IEEEauthorblockN{%
    Pouyan Fotouhi Tehrani\IEEEauthorrefmark{1},
    Raphael Hiesgen\IEEEauthorrefmark{2},
    Teresa Lübeck\IEEEauthorrefmark{2},
    Thomas C. Schmidt\IEEEauthorrefmark{2} and 
    Matthias W\"ahlisch\IEEEauthorrefmark{1}
  }
  \IEEEauthorblockA{%
    \IEEEauthorrefmark{1}TU Dresden, Dresden, Germany\\
Email: \{pouyan.tehrani,m.waehlisch\}@tu-dresden.de}
  \IEEEauthorblockA{%
    \IEEEauthorrefmark{2}HAW Hamburg, Hamburg, Germany\\
Email: \{raphael.hiesgen,teresa.luebeck,t.schmidt\}@haw-hamburg.de}
}

\maketitle

\tikzexternaldisable
\begin{tikzpicture}[overlay,remember picture]
  \node[draw, fill=gray!10, font={\scriptsize},anchor=south] at (.4\paperwidth,6cm) {\shortstack{If you cite this paper, please use the TMA reference:\\
	Pouyan Fotouhi Tehrani, Raphael Hiesgen, Teresa Lübeck, Thomas C. Schmidt and Matthias Wählisch. 2024.\\
  Do CAA, CT, and DANE Interlink in Certificate Deployments? A Web PKI Measurement Study.\\
  In \emph{Proceedings of the Network Traffic Measurement and Analysis Conference (TMA '24)}.
	IEEE, Piscataway, NJ, USA, 11 pages. \href{https://doi.org/10.23919/TMA62044.2024.10559089}{10.23919/TMA62044.2024.10559089}}};
\end{tikzpicture}
\tikzexternalenable

\begin{abstract}
Integrity and trust on the web build on X.509 certificates. Misuse or misissuance of these certificates threaten the Web PKI security model, which led to the development of several guarding techniques. In this paper, we study the DNS/DNSSEC records CAA and TLSA as well as CT logs from the perspective of the certificates in use. Our measurements comprise 4 million popular domains, for which we explore the existence and consistency of the different extensions. Our findings indicate that CAA is almost exclusively deployed in the absence of DNSSEC, while DNSSEC protected service names tend to not use the DNS for guarding certificates. Even though mainly deployed in a formally correct way, CAA CA-strings tend to not selectively separate CAs, and numerous domains hold certificates beyond the CAA semantic. TLSA records are repeatedly poorly maintained and occasionally occur without DNSSEC. 
\end{abstract}

\begin{IEEEkeywords}
DNS, DNSSEC, CAA, PKI, TLS, CT Logs
\end{IEEEkeywords}

\section{Introduction}
\label{sec:intro}

\begin{figure}
  \centering
  \includegraphics{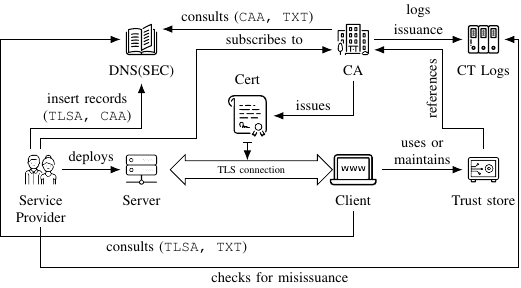}
  \caption{Overview of Web PKI entities and their relation to DNS(SEC).}
  \label{fig:webpki-ecosystem}
\end{figure}

Secure and authenticated transport is essential to the modern web.
Trust in the Web PKI is built on X.509 certificates and derives from accepted root certification authorities (CA).
Any CA, which is part of a valid trust chain, can issue a trustworthy certificate for any service. %
Millions of devices rely on this ecosystem for browsing, shopping, online banking, \etc
Given the immense reach of each CA, much effort has gone into securing the Web PKI (\autoref{fig:webpki-ecosystem})---most prominently DNS-based Authentication of Named Entities (DANE)~\cite{RFC-6698}, Certification Authority Authorization (CAA)~\cite{RFC-6844}, and Certificate Transparency (CT) logs~\cite{RFC-6962}.

DANE and CAA build on the Domain Name System (DNS), which  allows domain owners to store accessible information.
\textit{DNSSEC} adds an intrinsic chain of trust along the DNS hierarchy.
Different from the certificate chain-of-trust, only the entity that controls a domain can sign its records. %
Using DANE TLSA records, the DNS can provide additional information to verify X.509 certificates and even establish trust without relying on a CA.
In contrast, \textit{CAA} records allow domain owners to restrict which CAs are allowed to issue certificates for their domains.
Independent of the DNS, \textit{CT logs} are append-only databases that collect published certificate and make misissued certificates visible to the public.

All three standards---DANE, CAA, and CT logs---concern X.509 service certificates but use different methods, different publication channels, and are targeted at different audiences.
In this work, we want to learn about the use (and misuse) of these concurrent approaches to harden the Web PKI ecosystem. In a large measurement study, we collect data from 4M domains based on the Tranco top list, spanning DNS records, X.509 certificates, and CT log entries. We compare and analyze the different records in use with respect to their existence, content, and intended semantics, \ie correctness, consistency, and coherence. Our major findings read:

\begin{enumerate}
  \item  Nearly $9\%$ (357k) of all domains deploy at least one of either DNSSEC, DANE, or CAA; overlapping deployment is much smaller.
  \item CAA records are largely deployed correctly and consistent with certificate issuers ($>90\%$). Nevertheless, CA strings are not precisely defined; several strings match more than a single CA, one matches 21 CAs.
  \item CAA records are mainly deployed without DNSSEC; even some TLSA records lack DNSSEC protection.
  \item TLSA records are rare and often poorly maintained, some of which correspond to certificates that have been revoked or have expired since long. 
\end{enumerate}

The remainder of this paper is structured as follows.
\autoref{sec:background} introduces the Web PKI along with the relevant technologies for securing certificate deployments and related work. %
Our measurement method, data collection and its processing are explained in \autoref{sec:method}. %
\autoref{sec:deployment} reports on the deployment  of X.509 certificates and the DNS extension records and examine their correctness with respect to the actual certificates in use. 
\autoref{sec:consistency} combines information from DNS records, certificates, and logs to systematically explore consistency and coherence of the security information set.
We conclude in \autoref{sec:conclusion} with an outlook on future research.

\begin{table*}
  \center
  \begin{tabular}{llllll}
  \toprule
  Technology & Core Idea & Infrastructure & Responsible & DNSSEC & Target Audience \\
  \midrule
  CT Logs~\cite{RFC-6962,RFC-9162} & Auditable certificate issuance & CT Logs & CA & n/a & Subscribers (Domain Owners)\\
  DANE~\cite{RFC-6698} & Bind public keys to names & DNS & Domain Owner & mandatory & Relying Party (Clients)\\
  CAA~\cite{RFC-8659} & Constraint issuer of domain names  & DNS & Domain Owner & not mandatory & CA (Certificate Issuer)\\
  \bottomrule
  \end{tabular}
  \caption{CT Logs, DANE, and CAA offer three different approaches to securing the Web PKI.}
  \label{tbl:webpki:technologies}
\end{table*}

\section{Background and Related Work}
\label{sec:background}
The Web PKI ecosystem is the foundation for authentication on the Web using X.509 certificates at its core, \eg \cite{tossw-saawm-21}.
Since its inception, various extensions have been introduced to enhance its functionality or address its shortcomings.
In this section, we briefly introduce Web PKI and discuss how DNS CAA, DANE, and CT Logs address one of the biggest challenges: unconstrained and global certification authority.
Furthermore, we present prior work.

\subsection{Background}
A fundamental security challenge on the Internet is the trust in cryptographic keys used for authentication.
In a controlled environment, keys can be manually attributed to specific entities, but such an approach is less applicable in large-scale distributed systems such as the Internet.
We now discuss how Web PKI addresses this challenge, which shortcomings still exist, and which remedies have been proposed.

\paragraph{Web PKI}
Web PKI introduces \emph{Certification Authorities} (CA) to bind public keys to domain names (among other attributes) to form a certificate.
The authenticity of a certificate can be verified using its cryptographic signature.
In public key cryptography, a signature is generated by a private key (which is kept secret) and can be validated by the corresponding public key (which is published openly).
A relying party (RP), \ie a piece of software that decides whether a certificate is valid or not, would then trust a certificate that is signed by or can be traced back to a trusted CA.
On the Web, an RP is typically a browser, which maintains its own set of trusted CAs or Trust Anchors (TA) in a local \emph{trust store}.

Issuing a certificate correctly is the most important task of a CA.
Unfortunately, CAs are not restricted when issuing certificates.
They can create certificates for any name.
A compromised or malicious CA, then poses security risks for all entities that rely on security assurances provided by the Web PKI.
Those incidents occur in practice, such as the DigiNotar incident~\cite{agsbc-mahsa-17}.
To counter this threat, various solutions have been introduced, which we summarize in \autoref{tbl:webpki:technologies} and discuss in the following.

\paragraph{Certification Authority Authorization (CAA)}
A CAA record~\cite{RFC-6844} gives a domain name owner the ability to restrict issuance of certificates by defining which CAs are allowed to issue certificates for its name.
Such a \emph{constraint} is stored in the DNS using dedicated CAA resource records to describe restrictions for wildcard or fully qualified domain names (FQDN) for the namespace under control of the name owner.

A CAA record is composed of a flag, a tag, and a value.
Issuance constraints are defined by \texttt{issue} and \texttt{issuewild} tags.
The latter only constrains wildcard certificates while \texttt{issue} records concern both, wildcards and FQDNs, but are superseded by an \texttt{issuewild} record.
These two tags have a well-defined syntax.
When the syntax is violated, a certificate should not be issued.
To forbid issuance explicitly, an empty value (\texttt{";"}) can be used.

CAA records allow learning details about the CA itself.
DigiCert, for example, accepts \texttt{digicert.com} as well as \texttt{amazon.com} (among others) according to its Certification Practices Statement (CPS)~\cite{DIGICERT-CPS}.

CAA also enables CAs to report policy violations to the name owner based on information configured in the \texttt{iodef} tag.
The value of this tag can be an email address or a URL.
An example of a policy violation is when a certification request is submitted at a CA that does not satisfy the issuance constraints.

\paragraph{DNS-Based Authentication of Named Entities}
DANE~\cite{RFC-6698} allows the attestation that a public key is a valid key for a domain name.
To enable this attestation, a domain name owner stores the public key in specific records of the name under attestation.
For TLS-based services, DNS TLSA records signal RPs which (type) of certificate to expect from the server.
This can be an end entity (EE) certificate (\eg a leaf certificate) or a TA certificate (\eg from a CA).
A TLSA record can reference a certificate or its subject public key information (SPKI).
The reference is either to the full raw data (\eg hex formatted certificate) or its digest (\eg SHA256 hash).

DANE makes use of DNS Security Extensions (DNSSEC) which bring authentication and integrity assurance to the DNS~\cite{RFC-4033} and mitigate common DNS attacks such as cache poisoning.
DNS records are signed in DNSSEC, have limited validity, and must be signed again after expiration.
Without DNSSEC, DNS records are susceptible to spoofing and manipulation, thus defeating the purpose of DANE.

\paragraph{Certificate Transparency Logs}
CT logs~\cite{RFC-6962,RFC-9162} were introduced to enable public monitoring and auditing of issued certificates.
Although logging is not mandatory~\cite{bf-brimp-24}, major browsers such as Chrome and Safari only accept certificates that are logged in at least two compliant CT logs.
Due to their market shares, this forces CAs to comply or lose out on large shares of customers.

\paragraph{Target Audiences}
Each technology follows a different core idea and is targeted at different parties, see~\autoref{tbl:webpki:technologies}.
Whereas CAA records are meant for CAs \emph{before} a certificate is issued, CAs feed CT logs \emph{after} issuance to make certificates visible to domain owners and thus allow verification.
DANE is deployed by domain owners as well, but faced towards certificate consumers, \ie relying parties, signaling which (type of) certificate to expect or accept.

\subsection{Related Work}
\label{sub:relatedwork}

Efforts to secure the Web PKI have been ongoing for more than 10 years.
CAA records were standardized in 2013~\cite{RFC-6844} and updated in 2019~\cite{RFC-8659}.
DANE became a standard in 2012~\cite{RFC-6698}, and CT logs were standardized in 2013~\cite{RFC-6962} and updated to version 2 in 2021~\cite{RFC-9162}.
Since 2017 the CA/B forum requests that CAs validate CAA records.
To the best of our knowledge, this paper is the first study that comprehensively analyzes all these technologies together to better understand configurations in real deployments, including inconsistencies.

\begin{figure*}
    \includegraphics{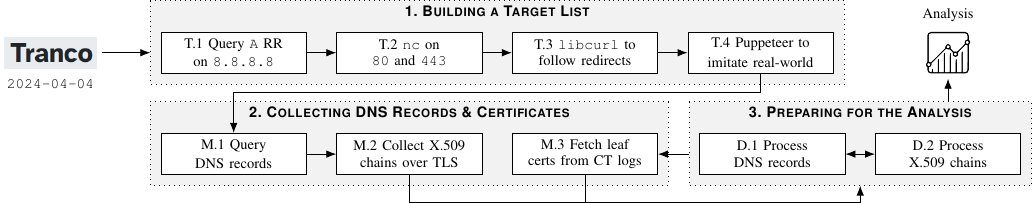}
    \caption{Simplified illustration of our toolchain for building a target list, collecting a dataset, and preparing data for analysis.}
    \label{fig:toolchain}
\end{figure*}

\paragraph{CAA}
In 2018, Scheitle~\etal~\cite{schgn-flcaa-18} presented the first analysis of the CAA ecosystem, actively measuring the behavior of selected CAAs as well as auditing the ecosystem.
They found that 3k of 95k domains in the Alex Top 1M list deployed CAA records.
At that time, most domain owners (89\%) configured a single CAA string, mainly letsencrypt.org (64\%), and did not allow issuing certificates for arbitrary names of a domain or subdomain (59\%).
CA strings that occurred infrequent included many invalid strings due to misspellings or owners using their own domain as CA string---indicating a lack of automation and understanding of how to configure CAA.
In a study focusing on nonfederal governments in the United States, Gebhard~\etal~\cite{gp-masdc-23} revealed that CAA is least deployed and grows slowly compared to the adoption of DNS records that strengthen identifying web servers (SPF, DMARC). %

\paragraph{Certificate Transparency}
In 2018, the same year Chrome made CT mandatory, Scheitle~\etal~\cite{sgnab-rctii-18} measured the adoption of CT and found an exponential increase of CT log entries.
They further found two more use cases of CT logs. \one CT logs can be used to identify phishing domains, \two~malicious actors monitor CT logs to search for new targets.
The latter was examined in a broader study in 2021~\cite{pnfkv-cvuct-23}, which confirmed the continued use of CT logs for target discovery. %
In a longitudinal study of TLS certificates gathered from active scanning and CT logs, Farhan~\etal~\cite{fc-eetc-23} found an improvement in share of valid certificates and key strengths, but also observe a centralization in the Web PKI. 
80\% of valid certificates are now signed by only 10~keys.

\paragraph{DANE}
Three years after DANE was published, Zhu~\etal~\cite{zwmh-mdta-22} found in 2015 that less than 1000 domains use DANE.
In 2020, DANE still did not gain widespread adoption in browsers.
While the email ecosystem saw a comparatively higher adoption rate, mismatches between TLSA records and certificates  and incorrect DNSSEC were still frequent~\cite{lgrkc-lcsde-20}.
Lee~\etal~\cite{lamrk-uhdms-22} observe that 94\% of SMTP servers still rely on the certificates issued by CAs when they deploy DANE.

All the prior studies provide an in-depth understanding of CAA, CT, or DANE deployments.
They focus, however, on each protection mechanism separately.
In this paper, we first provide an update of recent deployments and then close a gap by taking a comprehensive view on the deployments of all three technologies together.

\section{Method and Data Corpus}
\label{sec:method}

We collect a data corpus of 4M domains to examine the Web PKI ecosystem\footnote{Source code, raw data, and our analysis are available under \url{https://doi.org/10.5281/zenodo.11081271}.} using the processing pipeline in \autoref{fig:toolchain}.

\subsection{Building a Target List}

\paragraph{Query DNS (T.1)}
As input, we take the Tranco top list~\cite{lvtkj-tranco-2019} comprised of over 4.1M domain names ranked by popularity.
We try to resolve each name to an A record and only keep names with valid records.
$\approx533k$ entries did not resolve to an IP address while $1150$ timed out, likely due to trimmed domain names or dynamic DNS changes.

\paragraph{Check Ports (T.2) \& Transport (T.3)}
Next, we check ports 80 and 443 (TCP using \texttt{netcat}) and establish an HTTP or HTTPS connection over open ports with our own tool based on \texttt{libcurl}. 
We follow HTTP (\texttt{3xx}) and HTML (\texttt{http-equiv}) redirects. %
$145k$ domains allow neither HTTP nor HTTPS connections. %
The remaining names resolve to IP address and host a web server.

\paragraph{Browser (T.4)}
We then feed collected names %
into Puppeteer (T.4), a headless Chrome browser, and follow further redirects (including JavaScript). %
Our target list now contains $4M$ unique domain names (including intermediates).

\subsection{Collecting DNS Records \& Certificates}

\paragraph{CAA, DANE, and DNSSEC from DNS (M.1)}
We collect the following DNS resource records (RR) for each domain: SOA, A, CAA, TLSA, as well as \textit{contactemail} and \textit{contactphone} TXT records, see Appendix~\ref{appendix:dnsrecords}.
Additionally, we query CAA records for all parents of a given name by recursively removing the leftmost label up to the TLD, \eg for \texttt{www.example.co.uk} we query CAA records for the domain set \texttt{\{www.example.co.uk, example.com, co.uk\}}.
Queries set the \texttt{DO} flag to request DNSSEC records (if any) and have DNSSEC validated by the resolver.
For DANE, we only query TLSA records that are associated with TCP services on port 443 (HTTPS). %

We use Google recursive resolvers for all DNS queries due to their availability, reliability, and provision of a JSON API.

\paragraph{X.509 Certificates via TLS (M.2)}
For each domain name, we establish a TLS connection with the first retrieved IP address in our list, setting the domain name as the Server Name Indication (SNI). %
At this point, we do not validate certificates and we suppress TLS errors due to insecure cipher suites (OpenSSL security \emph{Level 0}).
By disregarding failures in TLS deployments, we can collect all available certificates to better understand shortcomings.

\paragraph{Leaf Certs from CT Logs (M.3)}
Domain owners might have applied for more certificates than the one deployed on their web server.
For example, Cloudflare issues \emph{backup certificates} for its customers to be able to swiftly replace keys in case of compromise~\cite{k-ibc-22}.
We collect certificates appended to CT logs for the subset of domain names that have either CAA or TLSA records,
using the open database provided by Sectigo under \texttt{crt.sh}.

\subsection{Preparing for the Analysis}
\label{sub:preparation}

\paragraph{Parsing and Validating CAA (D.1)}
We parse CAA \texttt{issue} and \texttt{issuewild} values according to the Augmented Backus-Naur Form (ABNF) and consider malformed entries as empty (semantically equivalent to \texttt{";"}).
Malformed entries forbid the issuance of a certificate~\cite{RFC-8659}. %
For records with \texttt{iodef} tags, we verify that the value is a URL with the correct scheme~(\texttt{mailto}, \texttt{http}, or \texttt{https}).

For records with the \texttt{issue} and \texttt{issuewild} tags, we devise an algorithm that matches a set of given CAA RRs to certification authorities, visualized in Appendix~\ref{appendix:caamatching}.
It uses the following classifications:

\begin{enumerate}
    \item \textbf{No CAA}: No applicable CAA RR was found.
    \item \textbf{Implicit Match}: No CAA RR constraints issuance.
    \item \textbf{Issuer Match}: At least one CAA RR matches cert issuer.
    \item \textbf{Issuer Mismatch}: No CAA RR matches the issuer.
    \item \textbf{Malf. Mismatch}: All CAA RR are malformed.
    \item \textbf{Empty Mismatch}: Only empty (\texttt{";"}) CAA RR.
\end{enumerate}

\noindent
This algorithm relies on a mapping from CAA issuer domain names (\ie values in \texttt{issue} and \texttt{issuewild}) to CA certificate properties.
Our mapping is based on the ``List of CAA Identifiers'' in the Common CA Database~\cite{common-ca-database}.
We enrich this list manually
\one based on in Certification Practice Statements (CPS) documented identifiers, and \two our own observations of undocumented string identifiers.
The CPS is usually linked in the certificate.
If this link is not valid, we manually identify the respective CA and browse its website to find the statement.

\begin{figure}
    \centering
    \includegraphics{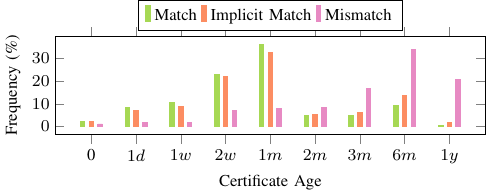}
    \caption{CAA matching states by age of certificates with matching subject name at the time of measurement. $75\%$ of certificates are younger than 3 months. CAA mismatches appear relatively more often in certificates older than 3~months.} %
    \label{fig:caa-time-diff}
\end{figure}

In contrast to X.509 certificates or DNSSEC signatures, CAA RRs do not carry a validity timestamp and are in practice only validated by the CA at the time of issuance.
This can cause a discrepancy between our observations at the time of measurement and what CAs observed when they issued a certificate, and thus leads to misclassifications.
To verify whether our measurement setup introduces such misclassifications, we calculate the difference between our probe time and \texttt{not before} timestamp carried in certificates for domains with CAA records, see \autoref{fig:caa-time-diff}.
Regardless of CAA matching state, our measurements occurred in 75\% within three months of the issuance.
Mismatches, however, occur much more frequently for certificates that are older.
This indicates that our setup is not prone to mismatches that are actually valid.

\paragraph{Parsing and Validating DANE (D.1)}
We validate that TLSA records are RFC-conformant using an open-source library~\cite{shuque-dane} but leave the verification of the DNSSEC integrity to the recursive resolver.

\paragraph{Parsing and Validating X.509 Certificates (D.2)}
We parse X.509 certificates with ZCrypto~\cite{zcrypto} and check three things.
\one The subject (alternative) names in the certificate should match the queried domain name, \ie the name is included as a SAN or covered by a wildcard SAN.
\two The certificate chain should be valid. %
We select the Mozilla Common CA Database~\cite{common-ca-database} as our trust store.
And \three Attached SCTs (if any) should be valid, \ie the SCT corresponds to the respective certificate and is signed by a trustworthy log.
We use Google's library~\cite{certificate-transparency-go} to verify signatures with keys extracted from the list of all complying~\cite{g-cctp-22} logs~\cite{known-ct-logs}. %
Here, we assume that log operators behave correctly %
and omit to check the logs directly.

\section{Configuration of X.509 Certificates, \newline DNS CAA, and DANE Records}
\label{sec:deployment}
\begin{figure}
  \centering
  \includegraphics[width=1.0\columnwidth]{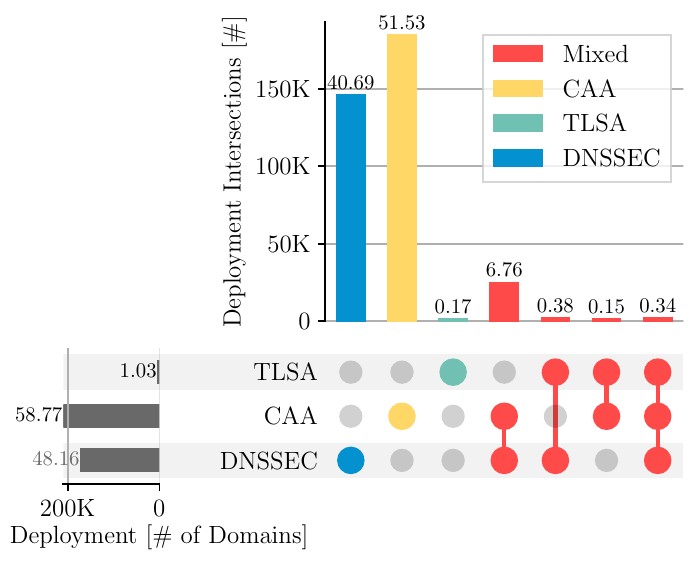}
  \caption{The UpSet plot contrasts the use of DNSSEC, CAA, and TLSA among domains that deploy at least one of them. CAA is most deployed ($\approx$52\%) followed by DNSSEC ($\approx$41\%) but they only co-occur in $<$7\% of domains. DANE (TLSA) is rarely deployed ($\approx$1\%).}
  \label{fig:deployment:upset}
\end{figure}
  
In this section, we focus on the deployment of X.509 certificates alongside CAA and DANE records within the DNS to better understand to which extent name owners care about correct configuration of each security extension.
In total, we observe 357k~unique domains (8.85\% of all names in the Tranco list) that deploy at least CAA, DANE, or DNSSEC.
We visualize the overlap of support for different technologies per name as an UpSet plot~\cite{lhsvp-uvis-14} in \autoref{fig:deployment:upset}.
The bar plot on the left shows how many domains fall into each category while the bar plot on top shows the size of exclusive intersections between the categories marked in the matrix below.
It is clearly visible that domain name owners do not focus on comprehensive security support.

\subsection{X.509 Certificates}
Roughly $97\%$ of collected certificates are valid (based on our trust store, see \autoref{sub:preparation}).
Validation failures stem from age ($47k$), untrusted signees ($\approx 15k$), and malformed certificates ($60k$).
All valid certificates were submitted to either 2 ($72\%$), 3 ($23\%$), or 4 ($5\%$) CT logs.
With $73\%$, most certs were submitted to logs operated by Google, two thirds were submitted to logs operated by Cloudflare, and about half to logs operated by DigiCert.

Two out of every three certificates were issued by one of four major CAs: Let's Encrypt with a total share of $52\%$, followed by Google Trust services ($16\%$), DigiCert ($4.5\%$), and Sectigo ($4.5\%$).
This does not include resellers, \eg ZeroSSL using Sectigo infrastructure.

In our observation, $7.2\%$ of unique domains point to hosts that provide certificates with mismatching subjects, \ie not matching the original domain name.
These are in part default (self-signed) server certificates or service provider certificates for parked domains.

\begin{table}
  \centering
  \begin{tabularx}{\columnwidth}{>{\ttfamily}X r r r r r}
    \toprule
    & \multicolumn{2}{c}{\tt issue} & \multicolumn{2}{c}{\tt issuewild} & \multirow{3.3}{*}{\shortstack{CA\\Count$^\dagger$}}\\
    \cmidrule(lr){2-3}
    \cmidrule(lr){4-5}
    \textnormal{CAA String} & Overall & Single & Overall & Single & \\
    \midrule
    letsencrypt.org         & 86.95\%  & 17.70\% & 85.28\%  & 6.43\% & 1 \\
    digicert.com            & 62.58\%  &  1.31\% & 77.94\%  &  1.36\% & 4 \\
    comodoca.com            & 45.68\%  &  0.39\% & 73.73\%  &  0.34\% & 12\\
    pki.goog                & 33.27\%  &  0.91\% & 51.10\%  &  0.06\% & 1\\
    globalsign.com          & 32.01\%  &  0.55\% & 31.13\%  &  0.63\% & 2\\
    sectigo.com             & 21.52\%  &  1.04\% & 25.37\%  &  1.79\% & 21\\
    \bottomrule
    \multicolumn{6}{l}{\shortstack[l]{$^\dagger$ Count of unique CAs (by Subject Organization) in our dataset that\\ match the CAA string in the first column}}
  \end{tabularx}
  \caption{The CA strings that appear in more than 10\% of the CAA records in our dataset as well as their relative occurrence as the \textit{only} CA string.}
  \label{tab:ca_string_shares}
\end{table}

\subsection{CAA Deployment}

$5.2\%$ of the $4M$ domains we scanned support CAA records.
Most ($4.55\%$) domains only specify constraints for CAs for fully qualified domain name or wildcard names~(respectively \texttt{issue} or \texttt{issuewild} records).
Around $0.01\%$ only provide information on reporting policy violations (\texttt{iodef} records).
$0.63\%$ of domains deploy both.

\paragraph{Reporting Policy Violations---\texttt{iodef} Records}
Several records that have been configured prevent contact because of misconfiguration.
About $3.79\%$ of \texttt{iodef} entries are invalid due to an invalid schema~\cite{RFC-8659} as neither \texttt{mailto} nor \texttt{http[s]} is present.
17 entries with unknown schemas still contain a colon, eight of these use a non-existent schema (\texttt{mailinto:}, \texttt{mail:}, etc.), six are typos with missing or switched letters, and three contain characters that break the formatting.
Among the remaining invalid entries, 924 are likely email addresses and 58 are HTTP~endpoints.
One record only contains a sequence of 27 numbers, which is unlikely a phone number because it is too long.
Among valid records, nearly all \texttt{iodef} records contain email addresses ($> 99\%$), 42 in total an HTTPS URL, and 16 domains have both.

We contacted domain owners with invalid iodef records.
Since many email addresses were associated with multiple domains, the 924 entries could be reduced to 504 distinct addresses.
Nearly 16\% of the mails could not be delivered, mostly because the mailboxes no longer exist.
While most did not respond, we also got kind responses and learned that at least one case was the result of false documentation of a hosting provider.
Others mentioned that their DNS settings were applied automatically by the service provider.

\paragraph{Granting Authorization to Issue (Wildcard) Domain Names---\texttt{issue} and \texttt{issuewild} Records}
Among $\approx$210k~domains with issuance constraints, $97\%$ have records with an \texttt{issue} tag and $57\%$ have records with an \texttt{issuewild} tag.
The overlap is $54\%$, which leaves $43\%$ that only have \texttt{issue} and $3\%$ that only have \texttt{issuewild}.
38~domains have malformed entries.

Our dataset shows that  compared to six years ago, name owners are more liberal when it comes to allowing multiple CAs to issue certificates.
In 2018, $89\%$ of domains only allowed a single CA to issue certificates~\cite{schgn-flcaa-18}.
Now, in 2024, $28.39\%$ domain owners allow four different CAs to certificate their names, $16.35\%$ 3 CAs, $16.26\%$ 5~CAs, and $10.04\%$ 2~CAs.
We even observe one domain with 59 CAA \texttt{issue} records and another one with 46 records.
Another 59 domains, all ending in \texttt{.ba.gov.br}, each have 17 CAA records.

\autoref{tab:ca_string_shares} shows the most commonly used CAA strings as well as the number of unique CAs (by Subject Organization) that match this string in our dataset.
Clearly, a domain owner grants permission to the operator of the CA~infrastructure rather than to a specific CA.
For example, ZeroSSL, an Austrian CA, accepts \texttt{sectigo.com}, because it uses Sectigo Infrastructure to issue certificates under its own brand.
At the same time, a CA might accept different CAA strings.
In our example, ZeroSSL also accepts \texttt{usertrust.com}, another brand of Sectigo.

\paragraph{Records with Non-Standard Tags}
We find 353 CAA records with tags not defined in RFC~\cite{RFC-8659}.
These tags can be categorized in three types: \one 252 $\times$ unrecognized tags as defined in CA/B Baseline Requirements~\cite[A.1.1.]{bf-brimp-24}, \two 50$\times$ misspellings such as an extra letter, and \three 51$\times$~malformed formats such as extra quotes.

\subsection{DANE Deployments}
\begin{figure}
  \includegraphics{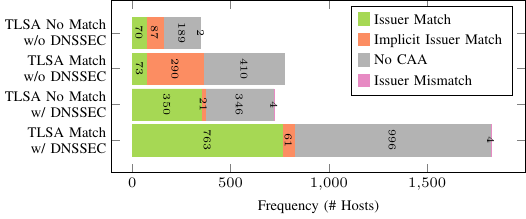}
  \caption{Total number of hosts delivering valid certificates with valid TLSA and CAA records divided by DNSSEC support, and TLSA and (selected) CAA matching status.}
  \label{fig:tlsa-caa-validity}
\end{figure}

About $0.1\%$ of all unique names that we queried are standard compliant.
98\% of these 3678 names provide a valid certificate.
The majority ($87\%$) define only end-entity constraints (DANE or PKIX EE), $10\%$ only trust anchor constraints, and the rest both.
Although DNSSEC is a requirement for DANE, one out of every third TLSA record set is delivered over 1126~insecure DNS sets.
\autoref{fig:tlsa-caa-validity} summarizes our findings.

It is noteworthy that only $70\%$ of names match their TLSA records as described below.
The case of mismatching TLSA records is discussed in depth in \autoref{sec:consistency} using data from CT logs.

\paragraph{Matching TLSA with Invalid Certificates}
About $1.5\%$ (53 names) define matching DANE-TA or DANE-EE constraints with invalid certificates (see \autoref{sec:method}).
More than half are expired and issued by Let's Encrypt and the remaining rest are self-signed certificates or leaf certificates by miscellaneous (partly not accredited) CAs.
More than $71\%$ of TLSA record sets here are secured by DNSSEC and $73\%$ have no associated CAA records.

\paragraph{Matching TLSA with Valid Certificates}
2553 entries provide TLSA records that match the provided certificate and the certificate is valid.
However, not all are secured by DNSSEC and only about $46\%$ of entries provides CAA records.
Only 190 names in this set have TLSA records that impose a limit on Web PKI certificates, while the rest are DANE-EE and DANE-TA constraints.

\section{Information Consistency}
\label{sec:consistency}

\subsection{CAA Records vs X.509 Certificates}
\label{sub:caa}

Based on the results of our CAA matching algorithm, described in \autoref{sub:preparation}, we examine the consistency between CAA records and the certificate issuers of the TLS certificates.
While the RFC~\cite{RFC-8659} mentions the role of an auditor (``Certificate Evaluators''), CAA records provide information for CAs at the time of certificate issuance and are not required to stay consistent.

\begin{figure}
  \centering
  \includegraphics{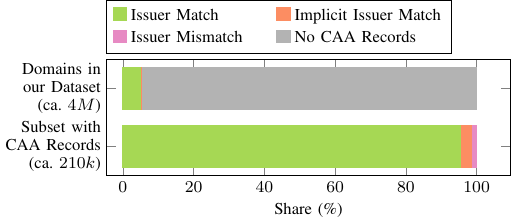}
  \caption{Most certificates with CAA deployment are consistent with their CAA records. We only observed an ``Issuer Mismatch'' in 2237 domains (1.07\% of CAA records).}
  \label{fig:deployment:caa:matching}
\end{figure}

\paragraph{CAA Validation Overview}
\autoref{fig:deployment:caa:matching} shows the result of our classification for two datasets based on CAA records with the \texttt{issue} tag.
The upper bar contains all $4M$ domains in our dataset while the lower bar only considers the subset with CAA records.
$3.8M$ domains ($94.8\%$) do not have CAA records.

$200k$ domains ($4.96\%$) have certificates consistent with their CAA records (``Issuer Match'').
These are $95.34\%$ of the nearly $210k$ domains with a CAA record.
Another $0.17\%$ ($3.22\%$) is classified as ``Implicit Issuer Match''.
These domains mostly (93\%) have CAA records with the \texttt{issuewild} tag but not with an \texttt{issue} tag while containing FQDN in their certificate, \ie the domain has the relevant resource records but only restricts the issuance for wildcard certificates.
$6\%$ only deploy \texttt{iodef} records. $<1\%$ only have records with unknown tags.

Nearly 91\% of domains with CAA records deploy the relevant RR, \ie the CAA record(s), themselves.
For roughly 9\% the direct parent domain has the relevant RR.
For about 200 domains ($\leq$1\%) the DNS hierarchy needs to be traversed further, up to 4 times, which occurred only once.

Only $0.06\%$ of all domains ($1.07\%$ with CAA records) have an ``Issuer Mismatch'', \ie the string in their CAA records does not match the issuer of their certificate.
We found two domains that deployed \textit{only} malformed CAA records.

\paragraph{Relevant CAA Records}
For most domains (92.4\%) a CAA record with the \texttt{issue} tag was the deciding record, \ie the CAA record that fit the domain we requested from the web server.
In 95\% of cases the issuer matches, 3.5\% have an implicit match, and the remaining 1.5\% mismatch.
In cases where the CAA record with the \texttt{issuewild} tag was relevant~(7.6\%), the share of domains with a matching issuer is even higher with 99.4\%.
0.6\% mismatch, and we do not observe any implicit matches.

\paragraph{Wildcard Certificates for Subdomains}
Next, we examine domains with \texttt{issuewild} CAA records and a wildcard name for their \textit{subdomains} in the certificate.
This is the case for 21\% of domains with CAA records in our dataset ($44k$).
For $99.5\%$ domains the issuer of the certificate is consistent with their CAA \texttt{issuewild} records.
216 ($0.49\%$) have a mismatch, \ie their certificates should not have been issued in this configuration.
17 ($0.04\%$) forbid issuance of a wildcard certificate with an empty CA string (but have an active wildcard certificate).

\paragraph{Partial CAA Matches}
A CAA \texttt{issue} record covers a FQDN and a wildcard domain in the absence of an CAA \texttt{issuewild} record.
However, a CAA \texttt{issuewild} record without an \texttt{issue} record would only restrict wildcard issuance.
If a certificate lists both FQDN and a wildcard domain, CAs should check for CAA \texttt{issue} and \texttt{issuewild} records.
A \textit{partial} match occurs when the issuer of a certificate with a FQDN and a wildcard matches either an \texttt{issue} or \texttt{issuewild} record, but not both.

For 60 domains, we can match the certificate issuer to the CAA \texttt{issue} record but observe a mismatch for the wildcard in the same certificate.
And for the reverse\textemdash the certificate issuer matches a CAA \texttt{issuewild} record, but not the \texttt{issue} record\textemdash we find 151 occurrences.
In no case is issuance restricted by an empty (`;') record.

It seems unlikely that domains, which change their CAA records in the time between re-issuance, change only parts of their records and that they do not choose to set an empty (`;') record.
As such, these are likely falsely issued certificates.

\subsection{CAA Records vs CT Logs}
In \autoref{sub:caa} we match the issuer of the certificates from web servers against their respective CAA records.
For all domain names with at least one CAA record, we query CT logs to fetch all other certificates bound to those names that are valid at the time of measurement. 
CT logs provide $1M$ certificates for about $191k$ unique domain names---reduced to $766k$ certificates after removing duplicates (\eg when a CA logs both precert and leaf).
To compare the results, we focus on domains with CAA records or mismatching server certificates.

\begin{figure}
  \centering
  \includegraphics{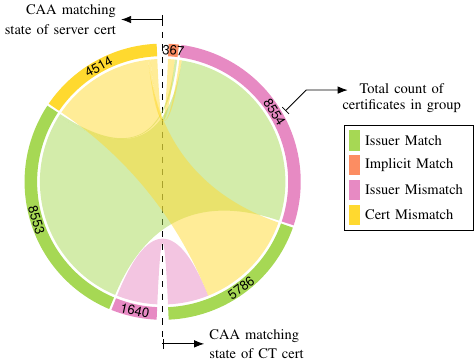}
  \caption{Distribution of inconsistencies between CAA matching state of server (left half-circle) and CT-logged certificates (right half-circle).}
  \label{fig:caa-orig-vs-ct-freq}
\end{figure}

In $98\%$ of cases the CAA matching state is consistent, \ie the certificates we retrieved from the web server and the certificates we retrieved from the CT logs have the same consistency with the domain's CAA records.
Among inconsistent cases, $\approx55\%$ of domains have servers that return a CAA-matching certificate while CT logs contain at least one other valid certificate that does not match the CAA constraints.
$30\%$ present a certificate that does not match the domain name (see \autoref{sec:deployment}) while we find a logged certificate that fulfills the CAA constrains.
\autoref{fig:caa-orig-vs-ct-freq} summarizes the findings.

\subsection{TLSA Records vs CT Logs}
DANE sets constraints on certificates provided by a service endpoint.
We take advantage of CT logs to find certificates that are not deployed, but match either TLSA records or the domain name of DANE-enabled services.
We explore \one the reasons for the existence of TLSA records that do not match the server certificate, \two the relationship between valid but not-deployed certificates and TLSA records, and \three compatibility of CAA records with certs referred to by DANE.

\begin{figure}%
  \centering
  \includegraphics{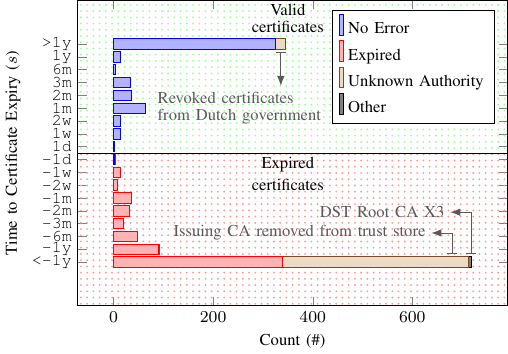}
  \caption{Certificates from CT logs that match TLSA records but are not deployed on the web server, grouped by their relative expiration time and annotated with their validation state.} %
  \label{fig:tlsa-cert-timediff-ct}
\end{figure}

\paragraph{TLSA Records that do not Match the Server Certificate}
We observe 1171 names with TLSA records that do not match the provided certificate.
Thus, DANE-enabled clients would consider these certificates invalid.
We query the \texttt{crt.sh} CT database for certificates that match these records and find 1494 certificates (note that a single domain can have multiple TLSA records).
\autoref{fig:tlsa-cert-timediff-ct} depicts how these are distributed with respect to their validity (\autoref{sub:preparation}) and %
relative expiration time.

The majority of mismatching TLSA records references certificates (leaf or CA) that are expired or have been removed from current trust stores.
An indication that TLSA records have not been updated to reflect changes of the CA or leaf certificate.
We also see TLSA records reference existing and valid certificates which are not deployed.

\paragraph{Undeployed Certs that Match TLSA Records}
For 945 domain names with TLSA records that fit their respective web server certificate we find 1002 more matching certificates in CT logs that are \emph{not} included in the certificate chain of the respective web servers.
The majority (784) are leaf certificates, and the rest (218) are from intermediate or root CAs.

All leaf certificates are either renewed or expired versions (same public key) of the certificate provided by the server.

\paragraph{Certificates Matching DANE-secured Domains}
For all domain names with TLSA records, we query CT logs for matching certificates (subject or SAN) and keep only valid ones (relative to the measurement time).
We only fetch leaf certificates and not the complete chain due to limitations in \texttt{crt.sh} database API.
Thus, we can only authenticate TLSA end-entity (PKIX-EE or DANE-EE) records against logged certificates.
A total of 10358 certificates match 3320 unique domain names.

\autoref{tab:tlsa-domains-authentication-iss} summarizes for how many certificates can be authenticated by the TLSA records.
Notable are cases where a TLSA record matches the server cert but not the CT cert and vice versa (rows \#3 to 6).
The majority of server-match/CT-mismatches (row \#3) are TLSA records that specify a certificate by its fingerprint (and not SPKI), so that even if the same key is renewed, it will not match.
We also observe Cloudflare backup certificates (of row \#4), where multiple certificates are valid simultaneously at the same time.

\subsection{TLSA Records vs CAA Records}

Most DANE-secured certificates collected from CT logs have the same CAA matching status as their corresponding server certificates.
The only notable exception is a set of domain names with matching CAA records and TLSA records referencing a valid certificate that does not match CAA (19 in total).
We also observe 19 cases where the server returned a certificate that does not match its domain name, but TLSA records reference certificates that match CAA constraints (9 already expired).

\begin{table}
  \centering
  \footnotesize
  \newcolumntype{Y}{>{\centering\arraybackslash}X}
  \caption{Number of CT logged certificates for all DANE-secured domain names grouped by authentication status.}
  \label{tab:tlsa-domains-authentication-iss}
  \begin{filecontents*}{./figures/data/tlsa-domains-authentication-iss.csv}
num,TLSA_AUTHENTICATED_ORIG,TLSA_AUTHENTICATED_CT,SAME_ISS,N
1,\cmark,\cmark,\cmark,959
2,\cmark,\cmark,\xmark,1
3,\cmark,\xmark,\cmark,2432
4,\cmark,\xmark,\xmark,294
5,\xmark,\cmark,\cmark,70
6,\xmark,\cmark,\xmark,15
7,\xmark,\xmark,\cmark,1777
8,\xmark,\xmark,\xmark,431
\end{filecontents*}
\pgfplotstabletypeset[
    col sep=comma,
    every nth row={1}{before row=\dhline},
    every nth row={4}{before row=\midrule},
    every nth row={6}{before row=\dhline},
    header=true,
    begin table={
      \begin{tabularx}{\columnwidth}
    },
    end table={\end{tabularx}},
    every head row/.style={ 
      output empty row,
      before row={%
          \toprule
          & \multicolumn{2}{c}{TLSA Record Authenticates:} & \multirow{3.3}{*}{\shortstack{Same Issuer\\(Server and CT)}} & \\
          \cmidrule{2-3} 
          \# & \shortstack{Server Cert} & \shortstack{CT Cert} & & Count \\
          \midrule
        }
    },
    every last row/.style={
        after row=\bottomrule
    },
    columns/TLSA_AUTHENTICATED_ORIG/.style={string type,column type=Y},
    columns/TLSA_AUTHENTICATED_CT/.style={string type,column type=Y},
    columns/SAME_ISS/.style={string type,column type=Y},
    columns/N/.style={column type={r}},
  ]{./figures/data/tlsa-domains-authentication-iss.csv}
\end{table}

\section{Discussion}
\label{sub:discussion}

In this section, we discuss our findings comprehensively and provide recommendations for name owners and CA~operators.

\paragraph{Lack of DNSSEC deployment defeats CAA purpose}
Missing DNSSEC deployment can be abused to trick CAs into misissuing domain-validated certificates~\cite{sww-difsd-19}.
While the CAA~standard does not require CAA records to be secured by DNSSEC, the use of DNSSEC is ``strongly RECOMMENDED''~\cite{RFC-8659}.
Yet, we observe a small overlap in the deployment as only $\approx7\%$ of CAA-enabled domains are also DNSSEC-secured, opening an unnecessary attack surface.

\paragraph{Reused strings in CAA \emph{issuer-domain-name} weaken security}
Not only do some CAs accept multiple strings, but some strings are accepted by multiple CAs.
We found one string accepted by up to 21 CAs, see~\autoref{tab:ca_string_shares}.
While these are likely resellers or managed PKIs, this nonetheless weakens the restrictiveness the system intends to implement.
Users are likely unaware of this when they choose their CA and create CAA records.
This conflicts with the goal of CAA records.

\paragraph{An authoritative source for CAA mappings can improve deployment and security}
We found many invalid strings clearly indicating that CAA~records are configured manually, even though supportive tooling (\eg {\tt\url{https://sslmate.com/caa/}}) exists.
A well-defined interface to discover legitimate CAA strings of a CA could help avoid typos, ease the setup, and increase the stability of the ecosystem.
Furthermore, the current specification and deployment model impacts auditors who must regularly scan the Certification Practice Statements~(CPS) of CAs to manually collect all valid CAA strings, which may lead to incorrect and incomplete data sets.

\paragraph{CT Logs can help CAA evaluators}
A CA only validates issuance constraints at the time of certification.
An evaluator, however, can validate CAA only after certificate issuance and might face CAA records other than what the CA observed.
This also explains why the number of CAA mismatches grows when the certificate age increases.
This time discrepancy can be reduced to a minimum by also acting as a CT Log Monitor and validate CAA records as soon as a certificate (or its precert) is logged.

\paragraph{DANE can enhance CAA}
The CAA standard~\cite{RFC-8659} claims that DANE records are relevant \emph{after} certificate issuance.
In fact, we found that DANE records are beneficial along the whole process and can even partly substitute CAA~features.
A TLSA TA record referencing root or intermediate certificates of a CA, for example, can be used to convey the same semantics as a CAA \texttt{issue} tag.
Furthermore, a ``to be signed certificate'' provided to the CA for signing already carries the public key and can be referenced by a TLSA record before certification.

\paragraph{Third party validation can help name owners}
Our correspondences with operators (\autoref{sec:deployment}) reveal that errors in CAA entries are partly due to incorrect instructions or automatic configurations by service operators (\eg hosting and DNS providers).
In addition, lack of feedback (using \texttt{iodef}) by issuing CAs in face of misconfigured or malformed CAA records makes it harder for name owners do detect such errors.
To address this, we have provided a public validation tool available on \texttt{\url{https://caa.secnow.net}}, which can validate CAA records, provide comprehensible explanations on effect of each CAA record, and detect errors and suggest solutions.

\section{Conclusion}
\label{sec:conclusion}

In this paper, we analyzed the Web PKI ecosystem with a focus on consistency between CAA, DANE, CT Logs, and X.509 certificates.
Ideally, each protection mechanism is correctly configured, and all agree on the set of legitimate certificates of a name.
Our results show different deployment behavior.
Most alarming, a small share of certificates should not have been issued since they contradict CAA records in place.
Furthermore, we found several configuration mistakes that could be prevented by better tooling and explicit best common practices.
We argue that there is room for improvement.
The CAA record structure itself is simple.
Clear information about the CAA strings and better support for configuring CAA records can help to increase adoption and help avoid common errors we identified.

We encourage domain owners to maintain their CAA records and keep them consistent with deployed certificates. 
Even more, we encourage extending the deployment to include DNSSEC and to consider TLSA records, as these techniques put web security on much firmer grounds.
Enabling automated evaluation of issued certificates and consider data of all possible protection mechanisms would be one step towards a consistent and healthy Web PKI.

\paragraph{Acknowledgments}
This work was supported in parts by the German Federal Ministry of Education and Research (BMBF) within the project PRIMEnet.

\vfill

\pagebreak

\bibliographystyle{IEEEtran}
\bibliography{internet, rfcs, visualization, own}

\appendices
\counterwithin{table}{section}
\renewcommand{\thetable}{\Alph{section}.\arabic{table}}

\counterwithin{figure}{section}
\renewcommand{\thefigure}{\Alph{section}.\arabic{figure}}

\section{Ethical Considerations}
\label{appendix:ethics}

Our measurements use common place actions (requesting DNS records, performing TLS handshakes) and were spread out to avoid hitting the rate limits or other protective measures.
The information we collect is public and explicitly designed to be checked by the relevant parties.
We neither reveal new security vulnerabilities nor do we single out any entity specifically. We individually notified all parties that were negatively impacted by misconfigurations or notable errors.

\section{X.509 Certificates}
\label{appendix:certificates}

\begin{figure}[h!]
    \centering
    \includegraphics{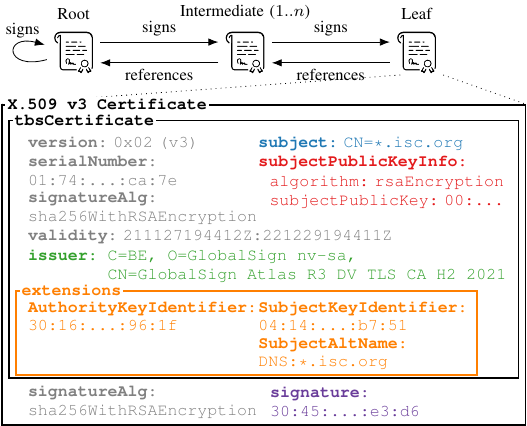}
    \caption{X.509 certificate chains and structure.}
    \label{fig:x509-overview}
\end{figure}

In the Web PKI X.509 build a chain of trust from a self-signed \textit{Root} certificate over one or more \textit{Intermediate} certificates to a \textit{Leaf}. As shown in \autoref{fig:x509-overview} the hierarchy is expressed through signatures from the root towards the leaf in one direction and references to the issuer in the other direction.

\section{Measurement Details}
\label{appendix:dnsrecords}
\autoref{tab:times} lists when each measurement task described in \autoref{sec:method} was performed and which tools we used.
\autoref{tab:dns-queries} provides additional information about the DNS queries when collecting X.509 Certificates via TLS (M.2).

\begin{table}[h!]
    \caption{Measurement dates and deployed tools for each task in our pipeline visualized in \autoref{fig:toolchain}.}
    \label{tab:times}
    \begin{tabularx}{\columnwidth}{c l >{\tt}c l}
        \toprule
        Task & Name & \textnormal{Date} & Tool \\
        \midrule
        T.1 & Query \texttt{AA} Records & \multirow{4.7}{*}{2024-04-07} & \texttt{dig} \\
        \dcline{1-2} \dcline{4-4}
        T.2 & Check ports 80/443 & & \texttt{nc} \\
        \dcline{1-2} \dcline{4-4}
        T.3 & Follow HTTP/HTML Redirects & & \texttt{libcurl}\\
        \dhline
        T.4 & Connect in Browser (Puppeteer) & 2024-04-12 & Puppeteer \\
        \midrule
        M.1 & Query DNS Records & \multirow{2.4}{*}{2024-04-12} & \multirow{2.4}{*}{JavaScript}\\
        \dcline{1-2}
        M.2 & Collect X.509 chains & & \\
        \dhline
        M.3 & Fetch certificate from CT logs & 2024-04-24 & Go \\
        \bottomrule
    \end{tabularx}
\end{table}

\begin{table}[h!]
    \caption{List of DNS resource records that we collect for a given domain name. Subdomain column denotes the prefix added to the domain name for the query; \texttt{@} denotes empty label.}
    \label{tab:dns-queries}
    \begin{tabularx}{\columnwidth}{l >{\tt}r l X}
        \toprule
        \# & \textnormal{Subdomain} & Type & Description \\
        \midrule
        1 & @ & SOA & Zone apex \\
        2 & @ & A & IP address(es) \\
        3 & @ & CAA & CAA RRs \\
        4 & \_443.\_tcp & TLSA & DANE \\
        5 & \_validation-contactemail & TXT & DV E-Mail contact\\
        6 & \_validation-contactphone & TXT & DV Phone contact\\
        \bottomrule
    \end{tabularx}
\end{table}

\section{CAA Matching Algorithm}
\label{appendix:caamatching}
\begin{figure}[t]
    \centering
    \includegraphics{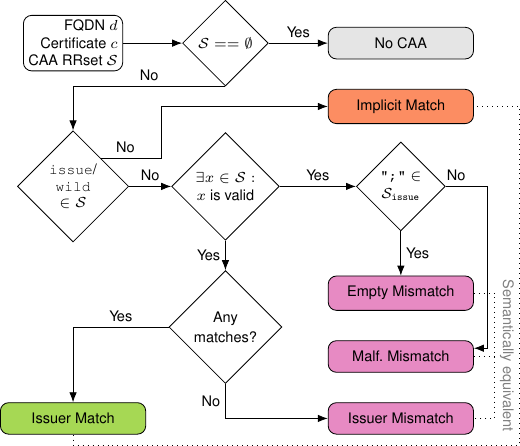}
    \caption{The algorithm we use to check the consistency between a certificate \textit{c} and relevant CAA resource records set \textit{S} for a given domain \textit{d}.}
    \label{fig:caa-flowchart}
\end{figure}

The consistency between CAA records and the certificate issuer can be classified into three main groups (colors in \autoref{fig:caa-flowchart}): {\sf\color{gray}No CAA} (relevant CAA RR set is empty), {\sf\color{Set2-D}Issuer Mismatch} (CAA RRs does not match the cert issuer), and {\sf\color{Set2-E}Issuer Match}.
Mismatching cases are further divided to denote if mismatch was caused by empty (\texttt{";"}) or malformed CAA \texttt{issue/issuewild} values.
{\sf\color{Set2-B}Implicit Match} is semantically equivalent with Issuer Match but specifies the case where relevant CAA RR set is not empty yet has no \texttt{issue} or \texttt{issuewild} records.

To match if a certificate matches an issuer denoted by a CAA \texttt{issue/issuewild} record, we have composed our own approach that maps a CA string identifier to a CA based on its subject name, organization name, or other features.
For example, for the following CAA RR:

\begin{verbatim}
example.com          CAA 0 issue "web.com"
\end{verbatim}

\noindent
we assert that the CA certificate carries `Network Solutions L.L.C.' as organization name in its subject field.

\label{lastpage}

\end{document}